\documentclass{article}
\usepackage[english]{babel}
\usepackage[a4paper,top=1.75cm,bottom=2.25cm,left=2.05cm,right=2.05cm]{geometry}
\usepackage{amsmath}
\usepackage{graphicx}
\usepackage[colorlinks=true, allcolors=blue]{hyperref}
\usepackage{xcolor}

\title{Regime-dependent reconstruction of structural connectivity: an \textit{in silico} study of spatially constrained modular neuronal networks}
\author
{Laia Barjuan$^{1,2}$, Hideaki Yamamoto$^{3,4}$, M. Ángeles Serrano$^{1,2,5}$ and Jordi Soriano$^{1,2,*}$\\
	\\
	\normalsize{$^{1}$Departament de F{\'\i}sica de la Mat\`eria Condensada, Universitat de Barcelona, 08028 Barcelona, Spain}\\
	\normalsize{$^{2}$Universitat de Barcelona Institute of Complex Systems (UBICS), Barcelona, Spain}\\
	\normalsize{$^{3}$Research Institute of Electrical Communication, Tohoku University, Sendai 980-8577, Japan}\\
	\normalsize{$^{4}$Graduate School of Information Sciences, Tohoku University, Sendai 980-8579, Japan}\\
	\normalsize{$^{5}$ICREA, Pg. Llu\'is Companys 23, E-08010 Barcelona, Spain}\\
	\normalsize{$^\ast$Corresponding author. E-mail: jordi.soriano@ub.edu}
}

\date{}

\begin{document}
\maketitle

\begin{abstract}

Inferring structural connectivity from neuronal activity is a major challenge in neuroscience, yet the conditions under which inferred interactions faithfully reflect anatomical connections remain unclear. Here, we investigate {\em in silico} how spatial constraints in a neuronal network and its dynamical regime jointly shape structural recoverability. We construct modular networks with varying average axonal lengths and simulate spontaneous activity using the Izhikevich model with synaptic depression across a broad range of excitatory strengths and background noise levels. Using Generalized Transfer Entropy, we infer effective connectivity and compare it with the underlying structural ground-truth at the levels of individual connections and global topology. We observe that the richness in the repertoire of activity patterns is highest in an intermediate regime between globally synchronized and noise-dominated activity, reflecting a balance between global excitation and stochasticity. Optimal structural reconstruction generally occurs under these same conditions, although this correspondence depends on network architecture and is weaker in highly modular networks. Together, these results show that structural recoverability depends jointly on network organization and dynamical regime. More broadly, our findings highlight the risks of treating inferred effective connectivity as a direct surrogate for structural connectivity.
\end{abstract}

\section{Introduction}

Understanding how neuronal network structure shapes collective activity, and how reliably that structure can be recovered from activity recordings, is a central challenge in neuroscience. \textit{In silico} neuronal networks provide a particularly useful framework for addressing this challenge: they offer simultaneous access to the complete structural connectivity and the resulting neuronal activity, while allowing systematic control over network architecture and dynamical parameters. These capabilities complement experimental recordings, in which large neuronal populations can be monitored but the underlying anatomical wiring typically remains inaccessible or only partially known.

Experimental techniques such as calcium imaging~\cite{grienberger2012imaging,grienberger2022two}, multielectrode arrays~\cite{williams2025vivo}, and implantable electronic devices~\cite{cho2022transparent} allow researchers to characterize collective neuronal behaviour both {\em in vitro}~\cite{soriano2023neuronal,yaron2025dissociated} and {\em in vivo}~\cite{stringer2024analysis,cheng2025large}. These measurements provide insight into the complex dynamics that emerge from interconnected neurons and how disruptions in network activity relate to changes in functional organization, particularly in the context of neurological disorders~\cite{Roudi2015Multi,yuste2024neuronal,Xiong2023Invivo}, but they do not directly reveal the structural connections underlying the observed dynamics. This limitation has motivated the development of analytical methods to infer structural connectivity from activity recordings~\cite{pastore2018identification,callegari2023modeling}. Statistical and information-theoretic approaches, including Transfer Entropy and its generalizations, have been shown to recover a meaningful fraction of the underlying structural connections from spike trains obtained through numerical simulations and calcium imaging data~\cite{schreiber2000measuring,stetter2012model,orlandi2014transfer}. However, the accuracy of such inference is sensitive to the analytical technique employed and the quality of the data, including acquisition rate and the recording modality~\cite{lutcke2013inference,de2018connectivity,wei2020comparison}. In addition, substantial effort continues to be directed at improving inference algorithms~\cite{po2025inferring}, but a complementary question remains: how does the dynamical regime of a network affect the accuracy with which its structural connectivity can be recovered?

This question arises because a given structural network can support different activity patterns depending on its dynamical regime~\cite{messe2014relating}. Consequently, changes in inferred connectivity may reflect changes in network dynamics even when the underlying anatomical wiring remains fixed. Numerical simulations by Stetter and coworkers~\cite{stetter2012model,orlandi2014transfer} demonstrated that network-wide bursts can strongly impair connectivity inference. During these highly coherent events, widespread neuronal activation makes direct interactions difficult to distinguish from indirect influences, producing reconstructed networks that approach random graphs. Activity outside network-wide bursts instead provides more reliable information about structural connectivity because it can more clearly reflect local, monosynaptic interactions. These findings motivate a systematic investigation of structural recovery across dynamical regimes.

This investigation, however, requires access to a broad range of activity patterns and sufficiently detailed knowledge of the underlying structural connectivity, a combination that remains difficult to achieve experimentally. Laboratory-grown neuronal cultures offer part of this combination. They are accessible to recording and manipulation, and their organization can be controlled using microscale engineering approaches~\cite{soriano2023neuronal,Xuu2023RecentProgres}. Topographical modulation~\cite{montala2022rich}, microfabrication-guided connectivity~\cite{gladkov2017design,qiu2025microtube,AMOS2026integrated}, and engineered modular architectures~\cite{yamamoto2018impact,MONMA2025106967} demonstrate how imposed architectural features shape collective activity. Whereas conventional cultures on flat, homogeneous substrates often exhibit network-wide bursts~\cite{wagenaar2006extremely,orlandi2013noise}, engineered cultures can display a broader repertoire of activity patterns, including locally confined bursts and network-wide synchronization~\cite{yamamoto2018impact,montala2022rich}. However, even in engineered cultures the complete structural connectivity remains inaccessible. Computational models inspired by these cultures remove this limitation, providing simultaneous access to the complete structural network and the resulting neuronal activity, while allowing systematic control over both network organization and dynamical parameters. They therefore provide a controlled framework for determining how the dynamical regime affects structural recovery.

Here, we use \textit{in silico} neuronal cultures to investigate the recoverability of structural connectivity from neuronal activity using Generalized Transfer Entropy. We construct networks with modular architectures and varying average axon lengths, and simulate spontaneous activity using the Izhikevich model with synaptic depression across a broad range of excitatory strengths and background noise levels. We compare the inferred effective connectivity with the known structural connectivity at the levels of both individual connections and global network topology. We find that the accuracy of structural reconstruction depends jointly on network organization and dynamical regime, and is highest in the regime of maximal activity complexity, except in highly modular networks. More broadly, our results provide a cautionary perspective on the common practice of treating inferred functional or effective connectivity as a surrogate for structural connectivity.

\section{Biologically-plausible structural networks}
\subsection{Construction of in silico neuronal networks}
\begin{figure}[t]
	\centering
	\includegraphics[width=\linewidth]{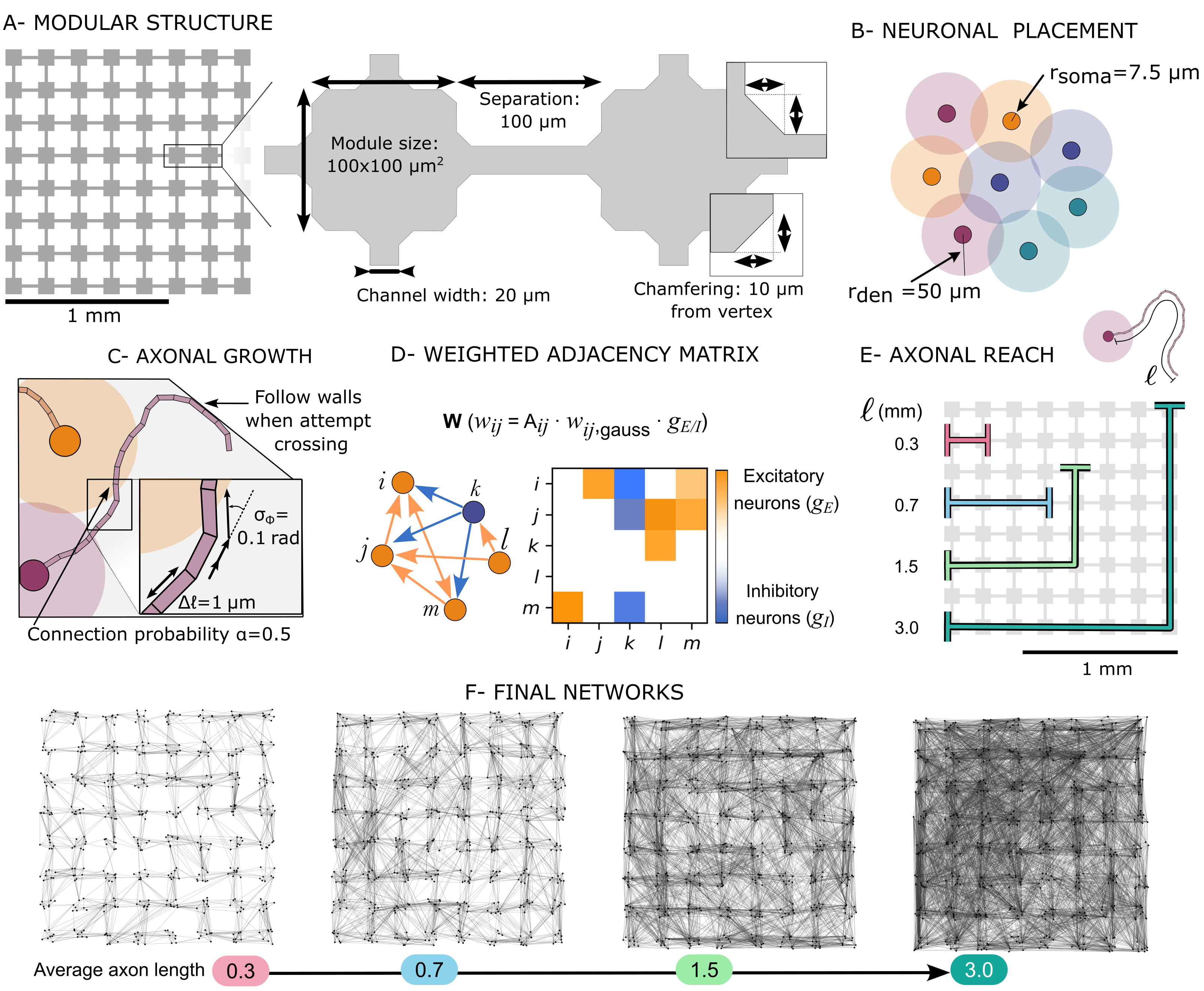}
	\caption{\textbf{Structural network construction.} A) Conceptual representation of the lattice-like modular pattern used to grow the neuronal culture. The inset on the right illustrates the dimensions of the modules and other topological features. B) Schematic of neuronal dimensions and placement within the culture. C) Illustration of the axonal growth mechanism, including axon-wall interactions. D) Graph and matrix representations of the structural connectivity. E) Illustration of the spatial extent within the culture corresponding to axon lengths $0.3, 0.7, 1.5$ and $3$ mm, in pink, blue, light green, and dark green, respectively. F) Networks generated for cultures with the different average axon lengths. Neurons are represented by black dots, and lines display all structural connections.}
	\label{fig:net_construction}
\end{figure}

Inspired by the experimental work of Yamamoto et al.~\cite{yamamoto2018impact,sono2026online}, we constructed \textit{in silico} models of neuronal cultures consisting of 64 interconnected square modules arranged in a lattice-like topology, each with a side length $m_l=100 \ \mu \textrm{m}$, and separated to its neighbours by a distance $m_s = 100 \ \mu \textrm{m}$. Connectivity between modules is restricted to channels of width $  c_w=20 \ \mu \textrm{m}$.  Figure~\ref{fig:net_construction}A shows the topographical pattern of the culture. We placed 8 neurons in each module ($\rho_n=800~\textrm{neurons/mm}^2$), which is of the same order as in experiments~\cite{sono2026online}, resulting in a total of $N=512$ neurons in the culture. We modelled the culture with $20\%$ inhibitory and $80\%$ excitatory neurons~\cite{soriano2008development}.

Broadly, the construction of the network takes place along three main steps. First, neurons are randomly placed inside each module such that overlapping somas are forbidden, and dendritic trees are represented as circles. Figure~\ref{fig:net_construction}B shows the schematic of neuronal dimensions. Then,  axons are grown according to the stochastic growth model presented in~\cite{orlandi2013noise}, where straight segments are concatenated until a total length $\ell$, independently drawn from a Rayleigh distribution with mean $\mu_{\ell}$. Whenever an axon attempts to grow in a direction that crosses a wall, it is forced to follow the wall instead, based on experimental evidence of contact guidance and boundary-following in micro-engineered environments~\cite{tanaka1995role,mahoney2005influence,li2005integration,francisco2007regulation}. Figure~\ref{fig:net_construction}C illustrates the axonal growth mechanism. Finally, directed connections are established probabilistically when an axon reaches the dendritic region of another neuron. Connection weights, $w_{ij,\textrm{gauss}}$, are drawn from a Gaussian distribution, and excitatory and inhibitory populations are assigned population-specific synaptic strengths to satisfy Dale's law~\cite{kandel2021principles}. The same parameter value is assigned to all neurons within each population. In particular, for inhibitory neurons the synaptic strength is $g_I=-12\,\mathrm{mV/ms}$, whereas for excitatory neurons the synaptic strength $g_E$ is varied across simulations between $2$ and $25$ mV/ms. All weighted connections are stored in the weighted connectivity matrix $\mathbf{W}=\{w_{ij}\}$. Each entry is the product of three terms: the adjacency indicator $A_{ij}$ (equal to $1$ if a connection $j \rightarrow i$ exists and $0$ otherwise), the Gaussian-distributed structural weight $w_{ij,\textrm{gauss}}$, and the synaptic strength $g_k$ of the population to which the presynaptic neuron $j$ belongs.  Figure~\ref{fig:net_construction}D illustrates both the graph and matrix representations of the weighted connectivity. Details of the network construction procedure and model parameters are provided in the Supplementary Material (SM).

This construction procedure defines connectivity within the physical geometry of the culture. The finite size and spatial arrangement of somas and dendritic territories determine where connections can potentially form, while axons explore this space through persistent trajectories of finite length and are guided by walls and micro-channels when encountering boundaries. Therefore, the topology of the resulting network emerges from the interaction between stochastic growth and the spatial constraints of the culture. Spatial embedding is a fundamental ingredient to consider when simulating neuronal cultures, as the resulting metric correlations and connectivity constraints can ultimately shape the network dynamics~\cite{orlandi2013noise,houben2025anisotropies,ludl2020impact}.

\begin{figure}[b!]
	\centering
	\includegraphics[width=\linewidth]{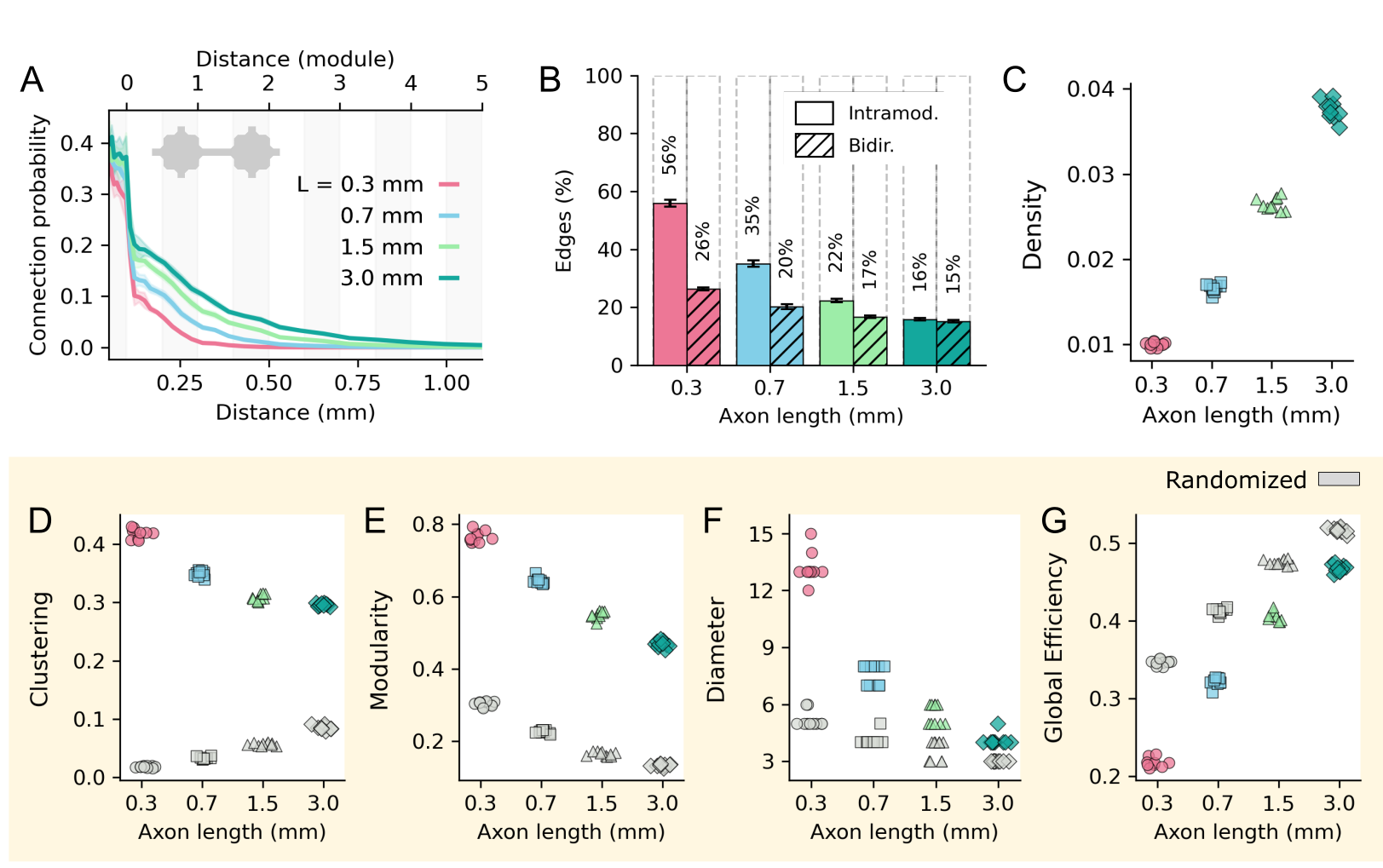}
	\caption{\textbf{Structural properties.} The results presented here are averaged over $10$ realizations. In A and B shaded regions or error bars indicate the standard deviation. In panels C-G, each symbol indicates an independent realization. A) Connection probability as a function of distance. The bottom axis shows the Euclidean distance, while the top axis indicates the corresponding approximate distance in modules, where $0$ denotes the module containing the source neuron. B) Fraction of network edges that are intramodular and bidirectional. The dashed gray line indicates the $100\%$. C) Density of connections. D) Average clustering coefficient, E) modularity F) diameter and G) global efficiency. In all panels, results for axon lengths $0.3, 0.7, 1.5$ and $3.0 \textrm{ mm}$ are shown in pink, blue, light green, and dark green, respectively. In panels D-G) the networks are analyzed as undirected and the results for degree-preserving random networks are displayed in gray.}
	\label{fig:struct}
\end{figure}

As previously introduced, the modularity of the substrate where neurons are grown, and consequently of the network of connections they establish, has been shown to increase the diversity of dynamical regimes exhibited by neuronal cultures, motivating our use of modular topographies. However, the specific value of modularity of the resulting structural networks depends on the parameters used during their construction. In our \textit{in silico} cultures, modularity can be tuned by varying either the width of the channels connecting neighbouring modules, $c_w$, or the average axon length, $\mu_{\ell}$. Smaller values of either parameter hinder the formation of intermodular connections, whereas larger values facilitate connections between modules. In the analyses presented here, we chose to vary the average axon length while keeping the channel width fixed. Specifically, we considered mean axon lengths of $0.3, 0.7, 1.5,$ and $3$ mm, which represent distinct regimes of spatial reach relative to the size of the culture, Fig.~\ref{fig:net_construction}E. In particular, an axon with a length of $0.3$ mm can reach at most two consecutive modules, one with a length of $0.7$ mm can span approximately four modules, a $1.5$ mm axon can traverse the entire length of the culture, and a $3$ mm axon can extend over roughly half of the culture's circumference. Figure~\ref{fig:net_construction}F shows representative structural networks generated for each average axon length.

\subsection{Network properties}\label{sec:netprops_micro}
We first characterize the structural properties of the networks generated by our biologically informed growth model. In Fig.~\ref{fig:struct}A we observe that, as expected in a spatially constrained network, the probability of connection is distance-dependent, being highest for neurons within the same module and dropping sharply across module boundaries. Increasing the average axon length extends the spatial range over which neurons can establish connections. For networks with an average axon length of $0.3$ mm, connections rarely extend beyond the first neighbouring module, whereas for $3$ mm axons, connections are still established between neurons separated by up to five modules. 

As a consequence, the proportion of intramodular connections decreases markedly with increasing axon length, Fig.~\ref{fig:struct}B. In particular, more than half of all connections are confined within the same module when the average axon length is $0.3 \textrm{ mm}$, whereas this percentage falls to $16\%$ for an average axon length of $3\textrm{ mm}$. Increasing axon length is also accompanied by a decrease in bidirectional connectivity, with the fraction of bidirectional connections falling from  $26\%$ for networks with $0.3$ mm axons to $15\%$ for those with $3 \textrm{ mm}$ axons. 

Beyond this redistribution of connections, longer axons also lead to an increase in the overall connection density. Although all generated networks remain highly sparse, connection density increases from approximately $0.01$ for networks constructed with shorter axons to $0.04$ for those constructed with the longest axons, Fig.~\ref{fig:struct}C.

We next examined the clustering coefficient, modularity, global efficiency and diameter, defined as the maximum shortest-path, of the generated networks. For the sake of simplicity, we treated the networks as undirected in the analyses. To assess whether the observed structural properties could be explained solely by the degree distribution, we compared the results of the original networks with those of degree-preserving randomized networks.

The average clustering coefficient ranges from approximately $0.45$ for networks generated with the shortest axons to below $0.3$ for those generated with the longest, Fig.~\ref{fig:struct}D. These values are already relatively high in absolute terms, but they are particularly striking when compared with randomized networks, whose average clustering coefficients are consistently below $0.1$.

A similar decrease is observed for modularity, Fig.~\ref{fig:struct}E, where networks generated with the shortest axons exhibit extremely high values, close to $0.8$, which progressively decrease to $0.5$ (still high) as the average axon length increases. The high modularity of the generated networks becomes particularly evident when compared with their randomized versions, as in all cases the corresponding randomized networks display modularity values approximately half those of the original networks.

The network diameter follows a related pattern, Fig.~\ref{fig:struct}F. The most modular networks (shortest axons) exhibit diameters of approximately $13$. Given that the cultures are arranged as an $8\times8$ grid of modules, any two modules are at most $15$ modules apart (including the endpoints). This indicates that, for these short-axon networks, the longest shortest path between two neurons traverses almost the entire culture one module at a time. As the average axon length increases, longer-range connections become more common and the network diameter decreases. Randomized networks have shorter diameters than the original networks, and this difference is less pronounced for networks with longer axons.

Finally, global efficiency displays the opposite trend, increasing from approximately $0.2$ to almost $0.5$ as axon length increases, Fig.~\ref{fig:struct}G. Nevertheless, in all cases, the global efficiency of the original networks remains lower than that of the corresponding degree-preserving randomized networks.

Summarizing, as the average axon length increases, the networks become progressively less modular and more integrated. This transition is reflected in a decrease in the clustering coefficient, modularity, and network diameter, accompanied by an increase in global efficiency. 

Altogether, the narrow shaded regions, small error bars and strong overlap of the data points corresponding to different realizations throughout Fig.~\ref{fig:struct} highlight the robustness and reproducibility of the generated structural networks.

\section{Neuronal network dynamics}

\subsection{Neuronal and synaptic dynamics: the Izhikevich model}

Having characterized the structural properties of the generated networks, we next investigate the dynamics that emerge on them. Individual neuronal dynamics were simulated using the Izhikevich model~\cite{izhikevich2003simple}, a quadratic integrate-and-fire model with adaptation that offers a good compromise between computational efficiency and biological realism, and that has been widely used in previous numerical studies of neuronal dynamics~\cite{orlandi2013noise,yamamoto2023modular,houben2025anisotropies}. We used a version of the model that also incorporates synaptic dynamics~\cite{orlandi2013noise}, which account for synaptic depression, such that tonically active neurons experience a reduction in synaptic transmission efficacy due to the depletion of available resources. 

Broadly, neuronal activity is described by the membrane potential $v_i(t)$ and the recovery variable $u_i(t)$, which evolve according to
\begin{equation}
	\frac{dv_i}{dt}
	= 0.04 v_i^2 + 5 v_i + 140 - u_i + I_i + \eta_i, \quad \textrm{and}
\end{equation}
\vspace{-0.35cm}
\begin{equation}
	\frac{du_i}{dt} = a \left( b v_i - u_i \right). 
\end{equation}
Here, $I_i(t)$ is the synaptic input current, which contains the synaptic inputs from other neurons through the weighted connectivity matrix $\mathbf{W}$, $\eta_i(t)$ is a Gaussian white noise term with intensity $\sigma_{\textrm{noise}}$ that triggers spontaneous neuronal activity~\cite{orlandi2013noise,hernandez2021noise}, and parameters $a$ and $b$ control the timescale of the membrane recovery variable and its coupling to the membrane potential. Excitatory and inhibitory neurons were described with identical parameters, differing solely in the sign and intensity of their synaptic strengths. A thorough description of the model and the parameters used in the simulations can be found in the SM.

\subsection{Dependence of the dynamical properties on model parameters}

Our goal is to characterize how the different dynamical regimes of the neuronal culture influence our capacity to extract structural information from the observed activity. To this end, we systematically varied some parameters of the model to explore a wide range of culture dynamics. Specifically, we explored different levels of stochastic input by varying the intensity of the additive Gaussian white noise, $\sigma_{\textrm{noise}}$ (mV$^2$/ms), from $1.5$ to $5$ in increments of $0.5$. In parallel, we varied the excitatory coupling strength, $g_E$ (mV/ms), considering values from $2$ to $12$ in increments of $2$, together with additional values $15, 20$ and $25$.  

\begin{figure}[t]
	\centering
	\includegraphics[width=\linewidth]{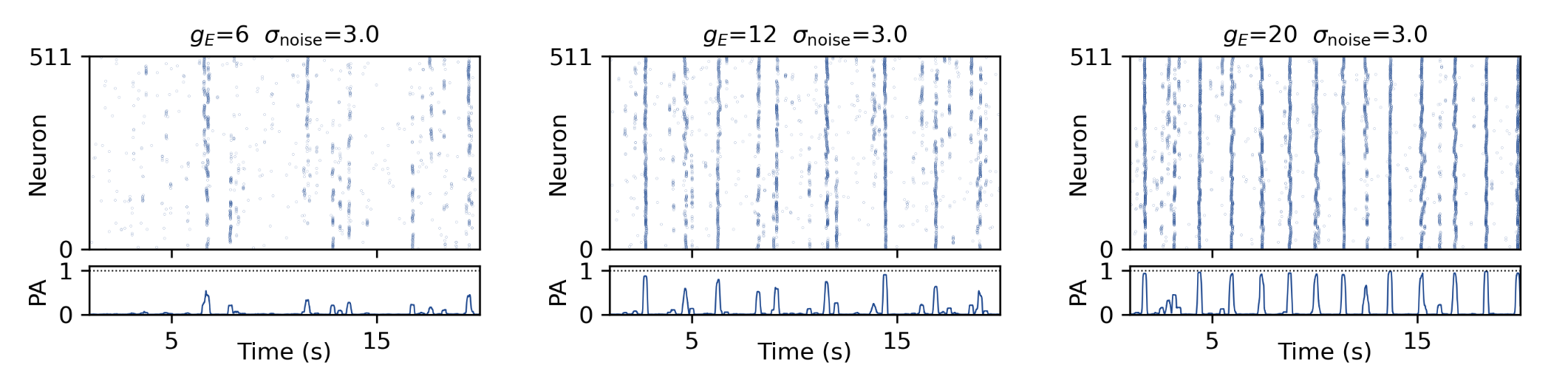}
	\caption{\textbf{Network activity for three parameter combinations and $\mu_{\ell}=0.7$ mm.} In each panel, the top plot shows the raster plot, where each blue dot represents an activation from the neuron in the y-axis, and the bottom plot shows the population activity over time. The x-axis represents time in seconds.}
	\label{fig:rasters}
\end{figure}

For each combination of parameters, neuronal activity was simulated for a total duration of $200$ seconds. Each simulation produced a sequence of spike events, specifying the firing time and identity of the neuron generating each spike. Figure~\ref{fig:rasters} illustrates the resulting neuronal activity as raster plots, showing the firing events of individual neurons over time. We also display the Population Activity (PA), which characterizes the collective activity of the neuronal population and is defined as the fraction of neurons that fire within successive sliding time windows, set to $200$ ms in our simulations. PA values close to $1$ indicate strong network bursting.

From these spike trains, we computed several measures to characterize the resulting dynamical regime. In particular, we examined the firing rate, the complexity of the population activity and the synchronization index. The firing rate is the average number of spikes emitted by a neuron per unit time and provides a measure of the overall activity level of the culture. The PA, as previously mentioned, represents the temporal evolution of the fraction of coactive neurons in the culture. Its complexity quantifies how uniformly the population activity explores its range of possible values, such that it equals $1$ when all activity levels are equally likely, indicating that the culture explores a rich repertoire of co-activation states, and is close to $0$ when the dynamics are concentrated around only a few activity levels, corresponding to a poor dynamical repertoire~\cite{yamamoto2018impact,montala2022rich}. Finally, the synchronization index, $\chi$~\cite{golomb1993dynamics,golomb2007neuronal}, quantifies the degree of coordinated activity across the neuronal population, and it is defined as the ratio between the temporal variance of the population activity and the average temporal variance of the individual neuronal activity traces. Large values of $\chi$ indicate that neurons tend to become active simultaneously, whereas values close to $0$ correspond to asynchronous activity. The mathematical definition of each of these measures is given in the SM.

\begin{figure}[h!]
	\centering
	\includegraphics[width=\linewidth]{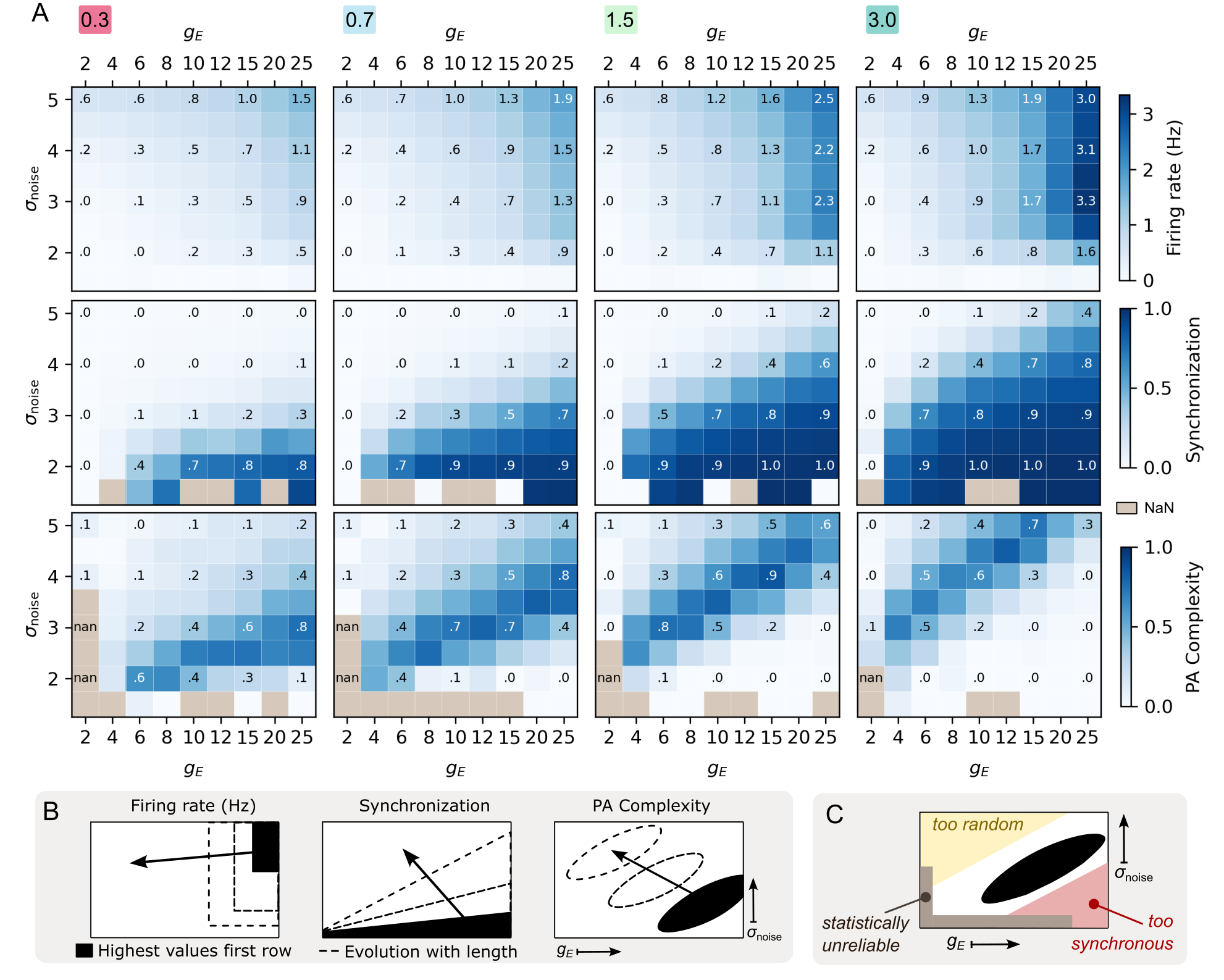}
	\caption{\textbf{Dynamical properties.}  A) Heatmaps of firing rate, synchronization index, and population activity (PA) complexity (top to bottom). Columns correspond to networks with average axon lengths of $0.3, 0.7, 1.5$ and $3.0$ mm (left to right). In each heatmap, the x-axis represents the excitatory strength (in mV/ms) and the y-axis the noise level (in mV$^2$/ms). Darker colours indicate higher values of the corresponding measure. Colour brown indicates a NaN value. Numerical values are displayed in every other row of the heatmaps. B) Schematic illustration of how the regions exhibiting the highest values evolve with increasing axon length within each column. The regions corresponding to the shortest average axon length are shown in black, while their evolution is indicated by dashed lines. C) Illustration of different dynamical regimes: the one dominated by random activity is highlighted in yellow, the highly synchronized regime in red, and when activity is statistically unreliable for analysis, in brown. The black area corresponds to maximum complexity.}
	\label{fig:dynamics_I}
\end{figure}

The firing rate exhibits a clear dependence on both noise and excitatory strength, as shown by the heatmaps in the first row of Fig.~\ref{fig:dynamics_I}A, where increasing either the noise level or the excitatory strength leads to higher firing rates, although the effect of the excitatory strength is more pronounced. The heatmaps reveal a clear increase in firing rate with average axon length, as observed from the left column to the right one. Networks constructed with longer axons exhibit higher firing rates across most of the parameter space, particularly at high excitatory strengths. This behaviour can be explained by the increased structural connectivity associated with longer axons, which facilitates the propagation of activity through the network. 

The synchronization index $\chi$ also depends on both dynamical parameters, as shown in the middle row of Fig.~\ref{fig:dynamics_I}A. When the noise level is extremely low ($\sigma_{\textrm{noise}}=1.5$ mV$^2$/ms), spontaneous activity becomes rare, and the cultures might not fire at all during the simulation. In these cases, the synchronization index is not defined, resulting in NaN values. When spontaneous activity does occur, neurons fire either as extremely isolated spikes or in highly synchronized bursts, yielding very low or very high values of the measure, respectively. Therefore, the synchronization index is highly dependent on the particular realization, preventing the emergence of a clear pattern in the bottom row of the heatmaps in this row of Fig.~\ref{fig:dynamics_I}A. At $\sigma_{\textrm{noise}}=2$ mV$^2$/ms, where cultures begin to fire more reliably, activity occurs predominantly through synchronized population events, with synchronization increasing as the excitatory strength grows. For moderate and high noise levels, on the other hand, networks with short axons do not achieve global synchronization, even for the largest values of the excitatory strength. As the average axon length increases, the region of high synchronization progressively expands across the parameter space. Networks with longer axons are able to sustain synchronized activity at higher noise levels and lower excitatory strengths, reflecting their enhanced structural connectivity. Overall, synchronization increases as the noise level decreases and the excitatory strength grows, with longer axons broadening the range of parameters over which synchronized dynamics can emerge.

The PA complexity exhibits a particularly interesting dependence on the dynamical parameters, as shown in the bottom row of Fig.~\ref{fig:dynamics_I}A. Unlike the firing rate or synchronization index, the PA complexity does not increase monotonically with either parameter. Instead, it reaches its highest values along a ridge that spans an intermediate region of the parameter space. For networks with the shortest axons, this ridge extends diagonally from low noise and low excitatory strength to moderate noise and high excitatory strength. Outside this region, the dynamics become less complex. In particular, combinations of low noise and high excitatory strength lead to highly synchronized activity (as evidenced by the synchronization heatmaps), reducing the diversity of population activity states and consequently lowering the PA complexity. At the opposite extreme, combinations of high noise and low excitatory strength produce dynamics dominated by random fluctuations, preventing the emergence of coordinated activity and likewise resulting in low complexity. Maximum PA complexity is therefore achieved at intermediate conditions, where the culture balances order and randomness. As the average axon length increases, this ridge of maximum complexity shifts toward progressively higher noise levels while preserving its diagonal orientation. This shift indicates that the increased structural connectivity associated with longer axons allows the culture to sustain rich dynamical behaviour under stronger stochastic input. 

For extremely low noise and excitatory strength, some NaN values appear in the PA complexity heatmaps. As for the synchronization index, at low noise levels, some of these values occur because the culture remains silent throughout the simulation. However, NaN values are more frequent for PA complexity and extend to low excitatory strength. This is because the measure depends on detecting peaks in the PA, which can occur either when the culture remains silent or when neuronal activity is too sparse to exceed the peak detection threshold. Thus, unlike the synchronization index, which can still be computed when isolated firing occurs, PA complexity is undefined not only in the absence of activity but also when activity is too sparse to produce detectable population-level peaks. Increasing the noise level raises the probability of neuronal firing, while longer axons enhance network connectivity and facilitate activity propagation, making coordinated bursts more likely and thereby reducing the occurrence of NaN values.

Figure~\ref{fig:dynamics_I}B schematically summarizes how the regions of highest firing rate, synchronization index, and PA complexity evolve as the average axon length increases. 

In summary, high noise levels combined with low excitatory strengths produce excessively random activity, whereas low noise levels combined with high excitatory strengths lead to highly synchronized dynamics. Between these two extremes lies an intermediate regime characterized by the highest PA complexity. Figure~\ref{fig:dynamics_I}C summarizes the distinct dynamical regimes observed and Fig.~SF1 in the SM shows illustrative raster plots for each of these regimes for an average axon length of $0.7$ mm. As mentioned above, the lowest excitatory strength and lowest noise level tested produce activity with poor statistical reliability. Consequently, in the following analyses, we exclude the parameter combinations with $\sigma_{\textrm{noise}}=1.5$ mV$^2$/ms and $g_E=2$ mV/ms.

\section{Effective network inference}
We now investigate the relationship between structural and effective connectivity inferred from the simulated activity. We estimate \textit{effective} rather than \textit{functional} connectivity because effective connectivity captures directed interactions and is less susceptible to spurious associations arising from widespread coactivation patterns, which are common in neuronal cultures. It is therefore better suited for assessing how accurately structural connectivity can be recovered from neuronal activity. For consistency with the terminology commonly used in the literature, we will nevertheless refer to the relationship between structural connectivity and activity-derived connectivity as the \textit{structure--function relationship}. Similarly, we will use the concept \textit{functional organization} to refer to the set of properties that characterize the inferred network.

\begin{figure}[h!]
	\centering
	\includegraphics[width=.7\linewidth]{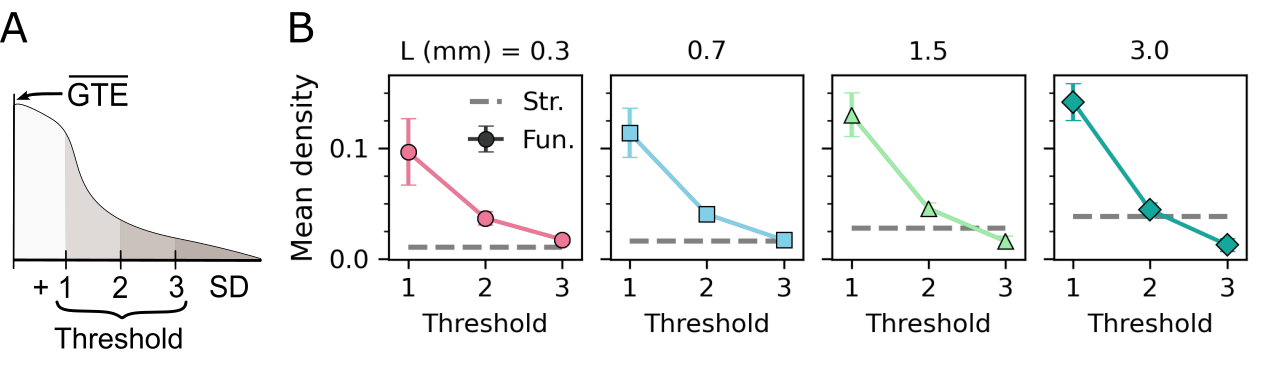}
	\caption{\textbf{Effective network density.} A) Illustration of the thresholds applied to the GTE distribution to infer effective connectivity. B) Connection density of the effective networks for each threshold and average axon length, averaged over all simulated dynamical parameter combinations. The dashed line indicates the ground-truth structural baseline.}
	\label{fig:strc_func_1}
\end{figure}

Throughout this work, effective connectivity is estimated using \textit{Generalized Transfer Entropy} (GTE)~\cite{stetter2012model}, which was specifically developed to infer directed interactions from neuronal spike trains. See the SM for a detailed description of the measure. The GTE algorithm assigns a score to every ordered pair of neurons, i.e., to every directed connection, resulting in an asymmetric, weighted effective connectivity matrix that is typically nearly fully connected. To identify the strongest inferred interactions, we standardized the GTE values as $z$-scores and applied three thresholds, $z=1$, $2$, and $3$, retaining only connections above each threshold. Figure~\ref{fig:strc_func_1}A schematically illustrates the effect of these thresholds on the distribution of GTE values.

An important consequence of this procedure is that the density of the effective connectivity network depends on the chosen threshold. Since this can affect the comparison between effective and structural connectivity, we first characterized how connection density varies across conditions. Figure~\ref{fig:strc_func_1}B shows the mean density of the effective network for each threshold and average axon length, averaged over all simulations with different values of noise and excitatory strength. As expected, increasing the threshold progressively reduces the density of the inferred networks. However, the relationship between effective and structural network density depends strongly on the average axon length. For networks with the shortest axons, even the highest threshold ($z=3$) produces effective networks that are consistently denser than the corresponding structural networks. As the average axon length increases, the density of the effective networks approaches that of the structural networks. For networks with an average axon length of $1.5$ and $3$ mm, thresholding at $z=3$ even produces effective networks that are, on average, sparser than the underlying structural networks.

\subsection{Link prediction assessment}

Our objective is to assess how accurately the inferred effective connectivity recovers the underlying structural network. This can be naturally formulated as a binary classification problem, where each possible directed connection between two neurons is classified as either present or absent. Comparing the inferred effective network with the ground-truth structural network link by link, and assuming the presence of a link in the effective network as a predicted structural connection, allows every possible connection to be assigned to one of four categories: a \textit{true positive} (TP), corresponding to a connection that is present in both the structural and inferred networks; a \textit{true negative} (TN), corresponding to a connection that is absent in both networks; a \textit{false positive} (FP), corresponding to a connection that is absent in the structural network but predicted to exist; or a \textit{false negative} (FN), corresponding to a connection that exists in the structural network but is not recovered by the inference method.

Several metrics have been proposed to quantify the performance of binary classifiers. A commonly used measure is the Receiver Operating Characteristic (ROC) curve, which represents the evolution of the true positive rate, or recall,
\begin{equation}
	\mathrm{TPR}=\frac{\mathrm{TP}}{\mathrm{TP}+\mathrm{FN}},
\end{equation}
as a function of the false positive rate,
\begin{equation}
	\mathrm{FPR}=\frac{\mathrm{FP}}{\mathrm{FP}+\mathrm{TN}}.
\end{equation}
The TPR quantifies the proportion of existing connections that are correctly recovered, while the FPR measures the proportion of non-existing connections that are incorrectly inferred as existing.

However, the ROC curve can provide an overly optimistic assessment when the positive class is rare, as in our highly sparse structural networks, where absent connections greatly outnumber existing ones. Because the FPR is normalized by the number of true negatives, even many false positives can result in a low FPR when the number of true negatives is large. Thus, for sparse classification problems, it is often more informative to complement the TPR with the \emph{precision}, defined as
\begin{equation}
	\mathrm{Precision}=
	\frac{\mathrm{TP}}{\mathrm{TP}+\mathrm{FP}},
\end{equation}
which measures the fraction of predicted connections that are actually correct.

\subsubsection*{True Positive Rate}
Regarding the reconstruction of our networks, the true positive rate (TPR) is highest for the most modular networks and progressively decreases as the average axon length grows (Fig.~SF2 in the SM). The high TPR values obtained for the most modular networks indicate that, under suitable dynamical conditions, the majority of structural connections can be successfully recovered from neuronal activity. As the networks become less modular, however, reconstruction becomes progressively more difficult. A plausible explanation is that the larger number of long-range connections creates additional indirect propagation pathways, making effective interactions less specific to direct structural links. Consequently, the information about the underlying structural connectivity becomes progressively diluted in the observed dynamics.

As expected, more restrictive thresholds $z$ yield lower TPR values, because fewer inferred connections are retained. For $z=1$, the maximum TPR decreases from approximately $0.9$ to about $0.65$ as axon length increases, whereas for $z=3$ these values reduce to $0.74$ and $0.26$, respectively. See Table~\ref{tab:TPR_Precision} for the values of maximum TPR for all thresholds and average axon lengths.

The TPR also depends on the dynamical regime. Overall, it increases with the excitatory strength, whereas its dependence on the noise level is non-monotonic. For intermediate excitatory strengths, moderate levels of noise maximize the TPR, while both lower and higher noise levels result in a poorer reconstruction. By contrast, for high excitatory strengths, the TPR increases steadily with noise.

\begin{table}[h!]
	\centering
	\caption{Summary of maximum values of TPR and precision.}
	\label{tab:TPR_Precision}
	\begin{tabular}{c| c| c c c | c c c }
		\hline
	     \multicolumn{2}{c| }{ } & \multicolumn{3}{c| }{TPR} & \multicolumn{3}{c}{Precision} \\ \hline
		 \multicolumn{2}{c |}{Threshold} & \textbf{1} & \textbf{2} & \textbf{3} &\textbf{1} & \textbf{2} & \textbf{3} \\ \hline
		Average & 0.3 & 0.90 & 0.84 & 0.74 & 0.16 & 0.30 & 0.43 \\
		axon & 0.7 & 0.87 & 0.76 & 0.62 & 0.19 & 0.34 & 0.47\\
		length & 1.5 & 0.77 & 0.60 & 0.43 & 0.23& 0.39 & 0.53 \\
		(mm) & 3.0 & 0.66 & 0.44 & 0.26 & 0.22 & 0.36 & 0.48 \\
		\hline
	\end{tabular}
\end{table}

\subsubsection*{Precision}
Although the TPR indicates that a large fraction of structural connections can be recovered, it does not account for the number of false positive predictions. As discussed above, this can lead to an overly optimistic assessment of reconstruction performance. 

The maximum values of precision are concentrated along a diagonal region of the parameter space, extending from low noise and low excitatory strength to high noise and high excitatory strength (Fig.~SF3 in the SM). Along this ridge, the precision increases progressively, reaching its highest values in the upper-right corner of the heatmaps, where both the noise level and the excitatory strength are largest. The location of this ridge indicates that reliable structural inference requires a balance between both dynamical parameters. 

Precision improves with the threshold $z$, since increasing the threshold eliminates many weak---and often spurious---inferred connections. Although this also discards some true positives, it removes an even larger proportion of false positives, leading to an increase in precision. The effect of the structural network is more modest: for a given threshold, precision is consistently slightly higher in the least modular networks. For $z=1$, the maximum precision increases from approximately $0.16$ to $0.22$ as axon length grows, whereas for $z=3$, these values increase to approximately $0.43$ and $0.50$, respectively. These values indicate that even under the most favorable conditions, only about half of the inferred connections correspond to true structural links, while for less restrictive thresholds fewer than one in four inferred connections are correct. See Table~\ref{tab:TPR_Precision} for the maximum precision for all thresholds and average axon lengths.

\subsubsection*{F1-score}

The TPR and the precision quantify complementary aspects of the agreement between the structural and effective connectivity. Although we have analysed these quantities separately, it is often useful to combine them into a single metric. For this purpose, we use the \textit{F1-score}, defined as the harmonic mean of precision and TPR:
\begin{equation}
	F1 =2 \ \frac{\mathrm{Precision}\times\mathrm{TPR}} {\mathrm{Precision}+\mathrm{TPR}}.
\end{equation}

This definition penalizes situations in which one of the two quantities is much smaller than the other. As a result, high F1-score values can only be achieved when TPR and precision are simultaneously high. The F1-score ranges from $0$ to $1$, which correspond to poor and perfect reconstruction, respectively. Poor reconstruction can arise either from a low recovery of structural connections, a large number of false positives, or a combination of both.

\begin{figure}[t]
	\centering
	\includegraphics[width=\linewidth]{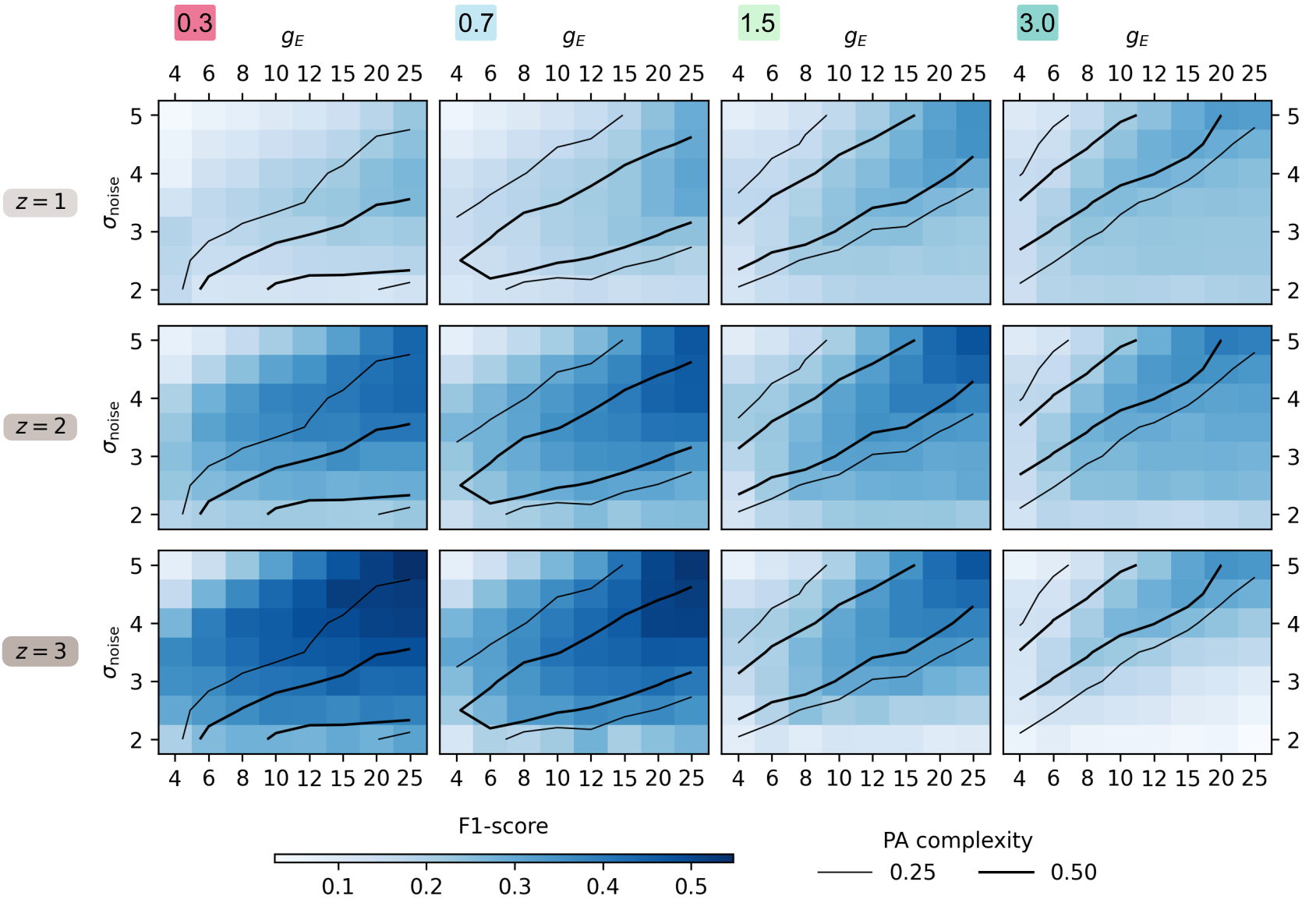}
	\caption{\textbf{Structure--function I.} Heatmaps of the F1-scores (top to bottom thresholds 1, 2, and 3). Columns correspond to networks with mean axon lengths 0.3, 0.7, 1.5, and 3.0 mm (left to right). In each plot, the x-axis represents the excitatory strength (in mV/ms) and the y-axis the noise level (in mV$^2$/ms). Darker colours indicate higher F1-scores. Black contour lines correspond to the PA complexity heatmaps, with thicker contours delineating regions of higher values.}
	\label{fig:strc_func_2}
\end{figure}

The heatmaps in Fig.~\ref{fig:strc_func_2} show the F1-score for all combinations of threshold, average axon length, and dynamical parameters (excitatory strength, $g_E$, and noise level, $\sigma_{\textrm{noise}}$). In all cases, the highest F1-scores are concentrated along a diagonal region of the parameter space, extending from low noise and low excitatory strength to high noise and high excitatory strength. Overall, the reconstruction performance improves for more restrictive thresholds and for more modular structural networks.

A particularly interesting observation is that the ridge of maximum F1-score approximately coincides with the region of maximum PA complexity. To facilitate the comparison, the contours of the PA complexity heatmaps are overlaid on the F1-score heatmaps. The correspondence is weak for the most modular networks, but becomes progressively stronger as the average axon length increases. This observation suggests that, whenever the structural networks are not extremely modular, the dynamical regimes that yield the best structure--function correspondence tend to coincide with those exhibiting the richest population dynamics.

\subsection{Characterization of effective network topology}

To further evaluate the alignment between the inferred effective networks and their structural baselines, we assessed how the graph properties of the inferred networks match or deviate from those of the original structural architectures across all combinations of thresholds, average axon lengths, and dynamical parameters. In particular, we analysed the same network metrics that we assessed before for the structural networks: density, average clustering coefficient, modularity, diameter, global efficiency, percentage of intramodular edges, and bidirectionality. For the calculation of clustering coefficient, modularity, diameter, and global efficiency, we symmetrized the matrices to enable the use of undirected network metrics.

\subsubsection*{The dynamical parameter landscape: excitation vs. noise}
For nearly all axon lengths and threshold configurations, within the dynamical parameter landscape $g_E \times \sigma_{\textrm{noise}}$, network metrics follow a continuous, diagonal phase gradient driven by a compensatory trade-off between excitation and noise. Broadly, the values increase as excitatory strength increases and noise decreases, reflecting a transition from noise-driven desynchronization to coordinated dynamics. Figure~\ref{fig:struc_func_3}A shows the direction of increasing modularity, which is representative of most metrics, except for global efficiency, which shows the inverse pattern. The connection density and the percentage of intramodular connections are exceptions to this rule too, and follow more diverse patterns within each individual $g_E \times \sigma_{\textrm{noise}}$ parameter space. These patterns are common to both, since the percentage of intramodular connections is directly related to connection density.       

\subsubsection*{Effects of thresholding}
The threshold $z$ acts predominantly as a global scaling factor rather than a driver of qualitative landscape changes. Global efficiency and connection density exhibit a highly coordinated global reduction as the threshold becomes more restrictive, whereas the intramodular edge percentage displays the inverse behaviour. In general, the spatial distribution of phases within individual $g_E \times \sigma_{\textrm{noise}}$ parameter spaces remains robustly invariant to the chosen threshold. However, for clustering, highly restrictive thresholds partially alter the pattern in the regions of the dynamical parameter space where density is very low, due to the mathematical difficulty of forming closed triplets at low edge densities.

\subsubsection*{Normalized Structure-Function Profiles}
\begin{figure}[t]
    \centering
    \includegraphics[width=\linewidth]{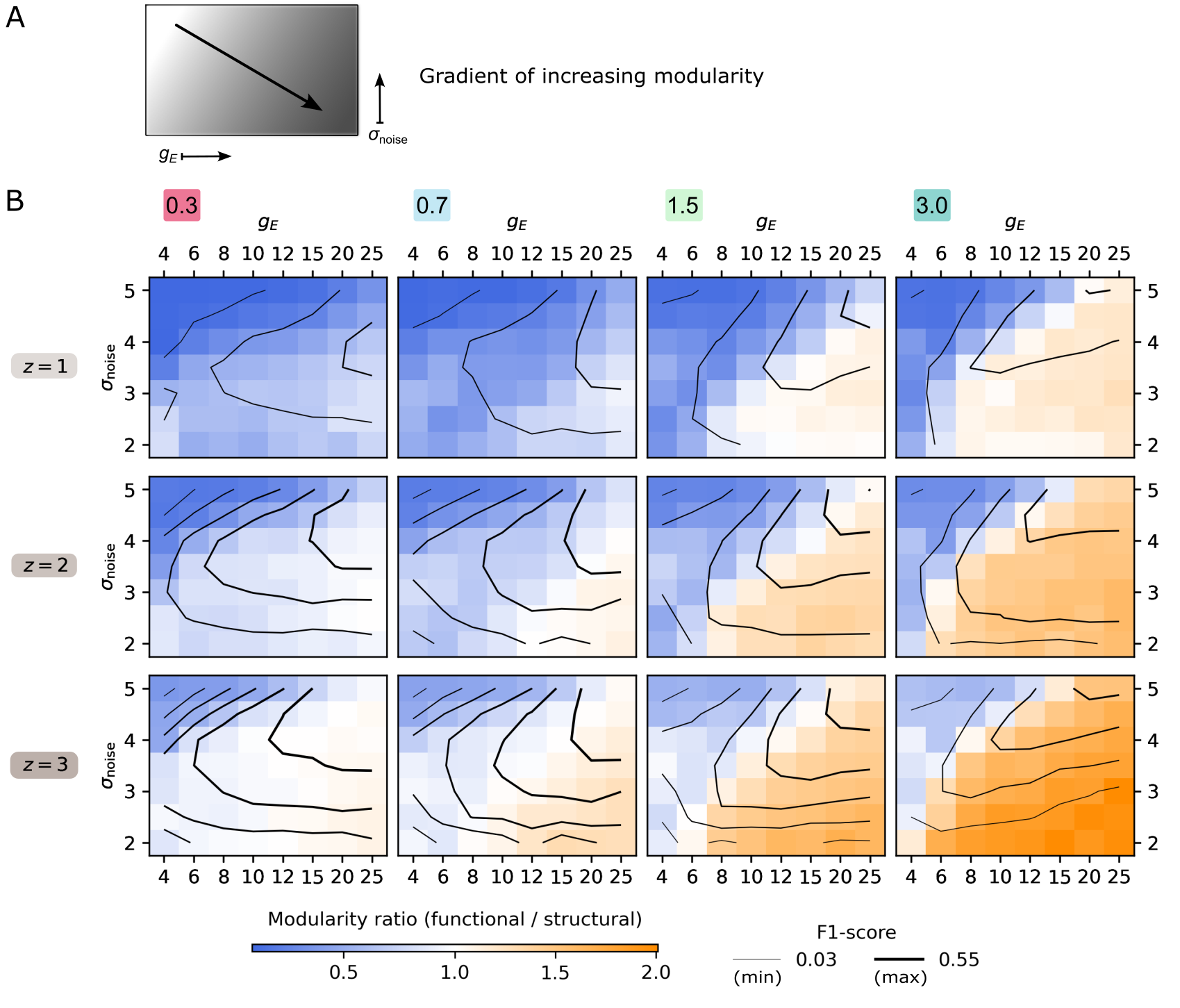}
    \caption{\textbf{Structure--function II.}  A) Illustration of the direction in which functional modularity increases within each plot. B) Heatmaps of the ratio between functional and structural modularity for thresholds $z=1, 2$ and $3$ (rows, top to bottom). Columns correspond to networks with mean axon lengths $0.3, 0.7, 1.5$ and $3.0$ mm (left to right). In each plot, the x-axis represents the excitatory strength (in mV/ms) and the y-axis the noise level (in mV$^2$/ms). Blue, white and orange indicate functional modularity lower, equal and larger than structural modularity, respectively. Black contour lines correspond to the F1-score heatmaps, with thicker contours delineating regions of higher F1-scores.	}
    \label{fig:struc_func_3}
\end{figure}

To identify the conditions under which the effective connectivity networks mirror the underlying anatomy, we normalized the derived functional metrics by their corresponding structural baseline. Consequently, a ratio of $1$ represents perfect structural equivalence, while values above or below $1$ imply functional hyper- or hypo-expression, respectively. The modularity results are shown in Fig.~\ref{fig:struc_func_3}B, while the corresponding figures for all other metrics can be found in the SM.

\begin{itemize}
	\item \textit{Modularity:}  Serves as a useful reference for assessing structural alignment, as it exhibits a combination of the trends of the other network metrics. When the average axon length is short, which translates to highly modular structural networks, reproducing the structural modularity would require extremely restrictive thresholds. For intermediate average axon lengths, the parameter space cleanly separates: in regions with high excitatory strength and low noise the system pushes modularity above the structural baseline, whereas for elevated noise and low excitatory strength the inferred network becomes less modular than the original one. 

    \vspace{0.1cm}
	\item \textit{Global Efficiency (GE):} Largely mirrors changes in network density. Consequently, it is lower or higher than the structural value whenever the inferred network is respectively sparser or denser than the underlying structural network. Whenever functional and structural densities are similar, GE reveals the influence of the dynamical regime. Highly noisy, weakly excitatory dynamics produce more random effective networks, leading to global efficiency values above the structural baseline. In contrast, low-noise, highly excitatory regimes generate effective networks with lower global efficiency than the structural network.
    \vspace{0.1cm}	
	\item \textit{Intramodular edges:} Strongly mirrors density trends but in the opposite way as GE: it lies above or below the structural baseline whenever the inferred network is respectively sparser or denser than the underlying structural network. An interesting exception arises in highly noisy, weakly excitatory regimes, where the inferred networks become so random that, despite being less dense than the structural network, they contain a smaller fraction of intramodular connections. Thus, in this regime, the increased randomness of the inferred connectivity outweighs the effect of density alone.
    \vspace{0.1cm}
	\item \textit{Clustering Coefficient:} Exhibits a unique decoupling from global network density. Across all axon lengths and thresholds, highly active, low-noise regimes force the network into dense, hyper-clustered functional triangles that substantially exceed the structural baseline. Increased noise and decreased excitatory strength sweep these local structures away, driving clustering below structural expectations regardless of network density. 
    \vspace{0.1cm}
	\item \textit{Diameter:} It does not have the smooth transition observed in other metrics. At low noise levels ($\sigma_{\textrm{noise}} \leq 2.5$ mV$^2$/ms), the effective networks frequently fragment into multiple connected components, preventing the diameter of the full network from being defined. This fragmentation is particularly pronounced for highly modular structural networks and for the most restrictive thresholds, where disconnected components are common. In contrast, for structural networks with intermediate modularity and more permissive thresholds, the fragmentation is generally limited to a few isolated neurons. 

    \vspace{0.1cm}
	\item \textit{Bidirectionality:} Systematically exceeds the structural baseline across all configurations. 
\end{itemize}
	
A particularly interesting observation is that, for metrics exhibiting both over- and underestimation, the transition between the two regimes broadly follows the direction of the ridge of maximum F1-score. In other words, the region where the topological properties of structural and effective networks are most similar aligns with the region where connection detection is most accurate. Together with the results of the previous section, this suggests that richer population dynamics tend to produce a more faithful representation of the underlying structural connectivity, both at the level of individual connections and global properties.

\section{Discussion}
Here, we studied whether and under what conditions, structural connectivity can be reliably recovered from neuronal activity. The results show that this correspondence is neither fixed nor determined solely by the inference method. Instead, structural recoverability depends jointly on the architecture of the underlying network and on the dynamical regime in which the network operates. In particular, the ability of activity to reveal structural connections is highest in an intermediate regime between global synchronization and noise-dominated activity, although this correspondence depends itself on the organization of the structural network.

First, we found that the dynamical landscape of the cultures is shaped by the underlying structural architecture. Across the parameter space, culture activity transitions from weakly coordinated, noise-dominated dynamics to increasingly synchronized activity as excitation increases and noise decreases. Between these two extremes lies a regime of maximal population-activity complexity, whose location shifts toward higher noise levels as the average axon length increases. Thus, the same balance between excitation and stochasticity does not produce the same collective dynamics in networks with different structural organization, indicating that structural and dynamical properties cannot be considered independently. This observation is consistent with studies exploring neuronal cultures grown on topographically modulated substrates, which have shown that substrate topography can influence the collective activity of the culture. In particular, spatial inhomogeneities in the form of equispaced parallel ridges were found to increase the dynamical repertoire of the culture, both {\em in vitro}~\cite{montala2022rich} and {\em in silico}~\cite{houben2025anisotropies}.

Structural organization also influences the recovery of structural connectivity from activity. The potential role of structural organization in shaping the agreement between structural and effective connectivity was suggested in~\cite{houben2025anisotropies}, where increased topographical anisotropy was associated with improved structural recovery. In our work, highly modular networks facilitate the recovery of existing connections but also produce slightly more false positives, while less modular networks make individual structural connections more difficult to recover but increase the reliability of the connections that are identified. These opposing trends show that structural organization affects different aspects of reconstruction in different ways and illustrate why a single measure of reconstruction performance can be misleading.

The F1-score, which balances recall and precision, identifies a diagonal region of high reconstruction performance extending from low-noise, weakly excitatory conditions toward high-noise, strongly excitatory conditions. This indicates that structural recoverability is associated with a balance between excitation and stochasticity rather than with a single optimal parameter combination. The ridge of maximum F1-score approximately follows the region of maximum population-activity complexity, but this correspondence is weak for the most modular networks and becomes progressively stronger as average axon length increases. Thus, complexity in the population activity should not be interpreted as a general predictor of structural recoverability: its relationship with reconstruction quality depends on the structural substrate on which the dynamics unfold. In highly modular networks, the activity can remain structurally constrained even when it is dynamically rich, and complexity alone is therefore insufficient to identify the regime in which individual connections are best recovered. 

The same interaction is also reflected at the level of global network organization. Modularity, clustering, global efficiency, and the fraction of intramodular connections exhibit systematic changes across the excitation--noise parameter space, and, for metrics that can be both over- and underestimated, their transition toward the structural values broadly follows the direction of the F1-score ridge. Thus, the dynamical conditions that favour recovery of individual structural connections also tend to produce inferred networks whose global organization more closely resembles that of the structural network. However, this agreement is not complete. In particular, bidirectionality is systematically overestimated across all configurations, which indicates that some distortions persist even when other network properties are close to their structural values. Moreover, low-noise regimes can lead to fragmentation of the inferred network, particularly in highly modular networks and under restrictive thresholds. These results emphasize that structure--function correspondence should be assessed using multiple complementary network metrics rather than a single topological descriptor.

The dependence on the inference threshold further emphasizes the trade-off between recall and precision. More permissive thresholds recover a larger fraction of structural connections but also retain many spurious interactions, resulting in high recall but low precision. More restrictive thresholds remove many false positives and therefore increase precision, but at the cost of discarding genuine structural connections. Structural recovery is therefore not only dependent on the dynamical and structural properties of the network, but also on how the continuous GTE interactions are converted into an effective network.

Our results are consistent with, and extend, the proposal by Stetter et al.~\cite{stetter2012model} that network-wide bursting episodes impair connectivity inference. While their results pointed to a distinction between highly synchronous bursting and more informative inter-burst activity, our systematic exploration of the dynamical parameter space reveals a broader continuum of reconstruction quality, with optimal recoverability occurring at an intermediate dynamical regime whose location in the parameter space depends on the structural organization of the network. Similarly, Barranca and Zhou~\cite{barranca2019compressive} found that connectivity could be reconstructed most accurately under dynamical conditions characterized by a balance between excitatory and inhibitory inputs. Although their model and reconstruction framework differ substantially from those considered here, their results support the broader principle that connectivity is not equally observable across dynamical regimes. In our model, the favourable dynamical conditions instead arise from a balance between excitatory strength and stochastic noise, rather than between excitation and inhibition.

Although our study focused on neuronal cultures simulated with the Izhikevich model and on effective connectivity inferred using Generalized Transfer Entropy, the qualitative conclusions are unlikely to depend exclusively on these specific choices. Future work should investigate whether similar optimal dynamical regimes emerge when using alternative inference methods, different neuronal models, different Izhikevich parameters, or even different strengths of both excitation and inhibition.

Taken together, our results demonstrate that the relationship between structural connectivity and activity-derived connectivity is conditional rather than deterministic: it emerges under specific dynamical conditions, and even then, agreement is partial and must be assessed across multiple complementary network metrics.

\enlargethispage{20pt}

\subsection*{Acknowledgments}
J.S. acknowledges support from the European Union Horizon 2020 research and innovation program under Grant No. 964877 (project NEU-CHiP), financial support from the Spanish Ministerio de Ciencia e Innovación under Project No. PID2022-137713NB-C22, funded by MCIU/AEI/10.13039/501100011033 and by ERDF/EU, and by the Generalitat de Catalunya under Project No. 2021-SGR-00450. L.B. acknowledges support from an FPU21/03183 grant from the Spanish Government. H.~Y. was supported by the MEXT Grant-in-Aid for Transformative Research Areas (A) ‘Multicellular Neurobiocomputing’ (24H02332), JSPS KAKENHI (22KK0177, 23H03489), and the Cooperative Research Project Program of the Research Institute of Electrical Communication (RIEC), Tohoku University. M.~A.~S and L.B. acknowledge support from grant PID2022-137505NB-C22 funded by MICIU/AEI/10.13039/501100011033 and by ERDF/EU, and Generalitat de Catalunya grant number 2021-SGR-00856.

\bibliography{Bibliography}

\end{document}